\documentclass[11pt,manuscript=article]{achemso}
\setkeys{acs}{maxauthors = 0}
\usepackage{graphics,color,amsmath}
\usepackage{subcaption}
\usepackage{comment}
\newcommand{\bm}[1]{{\mathbf{#1}}}

\title{Relative Binding Free Energy Calculations for Ligands with Diverse Scaffolds with the Alchemical Transfer Method}

\author{Solmaz Azimi}
\affiliation{Department of Chemistry, Brooklyn College of the City University of New York, New York, NY}
\altaffiliation{These authors contributed equally to this work}
\alsoaffiliation{Ph.D. Program in Biochemistry, The Graduate Center of the City University of New York, New York, NY}

\author{Sheenam Khuttan}
\altaffiliation{These authors contributed equally to this work}
\affiliation{Department of Chemistry, Brooklyn College of the City University of New York, New York, NY}
\alsoaffiliation{Ph.D. Program in Biochemistry, The Graduate Center of the City University of New York, New York, NY}

\author{Joe Z. Wu}
\affiliation{Department of Chemistry, Brooklyn College of the City University of New York, New York, NY}
\altaffiliation{These authors contributed equally to this work}
\alsoaffiliation{Ph.D. Program in Chemistry, The Graduate Center of the City University of New York, New York, NY}

\author{Rajat K. Pal}
\affiliation{Roivant Sciences, Boston, MA}

\author{Emilio Gallicchio}
\email{egallicchio@brooklyn.cuny.edu}
\affiliation{Department of Chemistry, Brooklyn College of the City University of New York, New York, NY}
\alsoaffiliation{Ph.D. Program in Chemistry, The Graduate Center of the City University of New York, New York, NY}
\alsoaffiliation{Ph.D. Program in Biochemistry, The Graduate Center of the City University of New York, New York, NY}

\begin{document}

\maketitle

\begin{abstract}
We present an extension of Alchemical Transfer Method (ATM) for the estimation of relative binding free energies of molecular complexes applicable to conventional as well as scaffold-hopping alchemical transformations. The method, named ATM-RBFE, implemented in the free and open-source OpenMM molecular simulation package, aims to provide a simpler and more generally applicable route to the calculation of relative binding free energies than is currently available. The method is based on sound statistical mechanics theory and a novel coordinate perturbation scheme designed to swap the positions of a pair of ligands such that one is transferred from the bulk solvent to the receptor binding site while the other moves simultaneously in the opposite direction. The calculation is conducted directly using a single solvent box prepared using conventional setup tools, without splitting of electrostatic and non-electrostatic transformations, and without pairwise soft-core potentials. ATM-RBFE is validated here against the absolute binding free energies of the SAMPL8 GDCC host-guest benchmark set and against a benchmark set of estrogen receptor $\alpha$ complexes. In each case the method yields self-consistent and converged relative binding free energy estimates in agreement with absolute binding free energies, reference literature values as well as experimental measurements.
\end{abstract}

\section{Introduction}
\label{sec:intro}

Relative Binding Free Energy (RBFE) calculations are a critical component of modern structure-based drug discovery efforts.\cite{Jorgensen2004,abel2017advancing,armacost2020novel} Generally, the goal of these calculations is to estimate the difference of the standard binding free energies of two molecular ligands to the same receptor more directly than taking the difference of the corresponding Absolute Binding Free Energies (ABFE). While a variety of approaches to obtain RBFEs have been proposed, most large-scale deployments to date employ a thermodynamic cycle in which one ligand is alchemically transformed into another ligand while both are bound to the  receptor and, separately, while both are in the solvent bulk. The difference of the free energy changes of these two transformations yields the RBFE of the ligand pair.\cite{wang2015accurate,cournia2017relative,cournia2020rigorous}

In RBFE methods, the transformations from one ligand into another are termed alchemical as they are based on progressively modifying the Hamiltonian of the system through a series of artificial chemical states that interpolate between the physical end states corresponding to the two complexes. In RBFE calculations, alchemical transformations are commonly carried out by parameter interpolation, in which the parameters of the energy function (partial charges, Lennard-Jones parameters, force constants, etc.) are progressively changed by means of an alchemical progress parameter $\lambda$.\cite{Mey2020Best,lee2020alchemical} Generally, the implementation of parameter interpolation methods require specific customization of the energy routines of the molecular dynamics engine, especially when soft-core pairwise potentials are used to treat end-point singularities.\cite{Steinbrecher2011,lee2020improved}

A variety of implementations of alchemical transformations are in common use. In single-topology implementations, a selected set of atoms of the first ligand are mapped to corresponding atoms of the second ligand. The first set of atoms is then transformed into the other by interpolating the force field parameters assigned to these atoms.\cite{liu2013lead} In the case of two ligands with different atom counts, the atoms of one ligand that do not correspond in the other ligand are converted to and from dummy atoms.\cite{fleck2021dummy} In dual-topology implementations by the means of congeneric pairing, a whole functional group of one ligand is typically converted to that of the second ligand by converting the first group to a dummy group and converting the second from a dummy group to the target group simultaneously.\cite{Mey2020Best} 

Single- and dual-topology approaches have distinct advantages and limitations that make them more or less preferable depending on the specific application. The correct treatment of dummy atoms in single-topology methods can be particularly subtle.\cite{fleck2021dummy} In dual-topology approaches, the mode of attachment of the dummy group to the ligand scaffold can also require care in order to avoid artifacts in the resulting free energies.\cite{zou2019blinded,Gallicchio2021binding} These and other implementation issues can be quite complex to address especially when considered in the context of hybrid approaches involving both single- and dual-topology transformations\cite{jiang2019computing}, as well as the more specialized separated-topology approaches.\cite{rocklin2013separated} In all of these methods, specialized system setup tools are employed to deal with the non-chemical molecular topologies to treat, for example, dummy atoms and provide ways to make sure that the atoms of the groups being transformed in dual topologies do not interact with each other. To avoid numerical instabilities, the alchemical transformations are also typically split into distinct simulation legs to separately treat van der Waals and electrostatic interactions.\cite{cournia2017relative,lee2020alchemical}

The setup of RBFE calculations based on two simulation systems---the receptor with the two ligands in the solvent and, separately, the ligands in the solvent---is further complicated by the preparation and simulation of multiple systems. RBFE implementations based on separate hydration and receptor-binding systems, for example, are not naturally suited for ligand pairs with different net charges.\cite{chen2018accurate} In these cases, the change of system charge during the alchemical transformation causes issues with the treatment of long-range electrostatic interactions and introduces artifacts in the free energy estimates unless complex correction factors are introduced.\cite{rocklin2013calculating} To overcome some of these challenges, coupled transformations have also been used in which the change in the net charge of the ligands is counterbalanced by the opposite change of charge of an ion to a solvent molecule or to another ion.\cite{dixit2001can,chen2013introducing} 

Scaffold hopping transformations that, unlike the simpler  terminal R-group modifications, include structural changes in the chemical topologies of the two ligands such as breaking and forming rings, reducing the size of rings, and modifying linker groups are particularly challenging with traditional RBFE approaches. Wang. et al.\cite{wang2017accurate} introduced a customized soft-bond stretch potential to address the numerical instabilities encountered when forming and breaking covalent bonds. More recently, Zou et al.\cite{zou2021scaffold} presented a method that uses a series of restraining potentials to address the same problem. 

As a result of the above discussed issues and other complexities associated with RBFE methods\cite{loeffler2018reproducibility} and despite the strong need in structure-based drug discovery, software tools in this arena remain of limited availability to many practitioners in the field. This deficiency is particularly felt in academic settings where commercial solutions might be prohibitively expensive or not sufficiently customizable, and the substantial amount of human resources and the specialized expertise required to develop and deploy RBFE solutions anew are not available. As an example, there are only a handful of open-source academic codes, at various stages of development, that are claimed to support scaffold-hopping RBFE transformations.\cite{jespers2019qligfep,vilseck2019overcoming,Rinikersubmitted}

In an attempt to address some of the aforementioned challenges, in this work we present a streamlined RBFE protocol based on the Alchemical Transfer Method (ATM).\cite{wu2021alchemical} The method aims to provide a simpler and more generally applicable route to relative binding free energy calculations than what is currently available. Like the parent ATM Absolute Binding Free Energy (ABFE) approach and unlike conventional parameter interpolation-based methods, the ATM-RBFE protocol is based on a straightforward coordinate transformation perturbation. As described here in this method, and similar to the double-system/single-box method\cite{gapsys2015calculation,macchiagodena2020virtual} for ABFE calculations, the receptor and the two ligands are initially placed in a single solvated simulation box in such a way that one ligand is bound to the receptor and the other is placed in an arbitrary position in the solvent bulk. Molecular dynamics simulations are then conducted with a $\lambda$-dependent perturbation alchemical potential energy function based on swapping the positions of the two ligands.

The ATM RBFE protocol proposed here has the following notable features. 
\begin{itemize}
    \item It is applicable to simple terminal R-group transformations, as well as to scaffold-hopping transformations to connect ligand pairs that do not share the same topology. 
    \item It does not require splitting the alchemical transformations into electrostatic and non-electrostatic steps, and it does not require the implementation of soft-core pair potentials. 
    \item It does not require modifications of the core energy routines of the molecular dynamics engine. It uses only the energies and forces returned by the unmodified existing routines used for molecular dynamics time propagation.  
    \item By design, it is applicable with any potential energy function, including conventional fixed-charge molecular mechanics force fields with and without long-range electrostatics, as well as many-body potentials, such as polarizable,\cite{harger2017tinker,panel2018accurate} quantum-mechanical,\cite{beierlein2011simple,lodola2012increasing,hudson2019use} machine learning potentials,\cite{smith2019approaching,rufa2020towards} implicit solvation models,\cite{Zhang2017Gaussian} and coarse-grained potentials,\cite{spiriti2019middle} which are generally poorly supported by free energy alchemical protocols.
    \item The molecular simulation system contains only chemically valid topologies without dummy atoms and is prepared with conventional tools. 
    \item Because perturbation energies are $\lambda$-independent and can be evaluated algebraically rather than by rescoring trajectories by calling the energy routines of the MD engine, the method is easily implemented in conjunction with advanced conformational sampling algorithms, such as Hamiltonian replica exchange,\cite{gallicchio2015asynchronous} and with multi-state free energy analysis tools.\cite{Shirts2008a,Tan2012}
    \item It is amenable to simple, compact and self-contained software implementations. The method presented here is implemented into a freely available and open-source plugin of the popular OpenMM library for molecular simulations.\cite{eastman2017openmm} 
\end{itemize}

The work is organized as follows. We first derive the formulation of the ABFE and RBFE ATM protocols from a well established statistical mechanics theory of non-covalent bimolecular binding. Here we also introduce ligand alignment restraints used to enhance the convergence of the calculations for similar ligand pairs and we show mathematically that they yield unbiased binding free energy estimates. We then present the application of the ATM RBFE method to a set of host-guest complexes and a set of protein-ligand complexes. Specifically, we compute all pairwise relative binding free energies for the SAMPL8 GDCC host-guest sets and for four proposed inhibitors of the estrogen receptor $\alpha$ nuclear receptor. Despite the fact that a majority of the ligand pairs considered in this work are related by challenging scaffold-hopping transformations, we show that the ATM RBFE protocol yields self-consistent relative binding free estimates that are in good agreement  with not only absolute binding free energy estimates, but also consistent with literature reference values and with experimental measurements. This work paves the way to streamlined and more readily available RBFE estimation tools in structure-based drug discovery and other fields.

\section{Theory and Methods}
\label{sec:theoryandmethods}

\subsection{The Alchemical Transfer Method}
\label{sec:ATM}

We review here the theory underpinning the Alchemical Transfer Method (ATM).  The presentation below tries to be mathematically rigorous, however simplified notation is often employed to unclutter the equations. For example, in intermediate formulas, we often omit limits of integration and Jacobian factors for curvilinear coordinates when they do not affect the form and interpretation of the final result. To shorten the expressions of the partition functions, the degrees of freedom of the solvent are often not indicated and included by means of the solvent potential of mean force $\Psi$.  

\subsection{Absolute Binding Free Energy Formalism}
\label{sec:ABFEwithATM}

\begin{figure}[bt]
\begin{centering}
  \includegraphics[scale = 0.5]{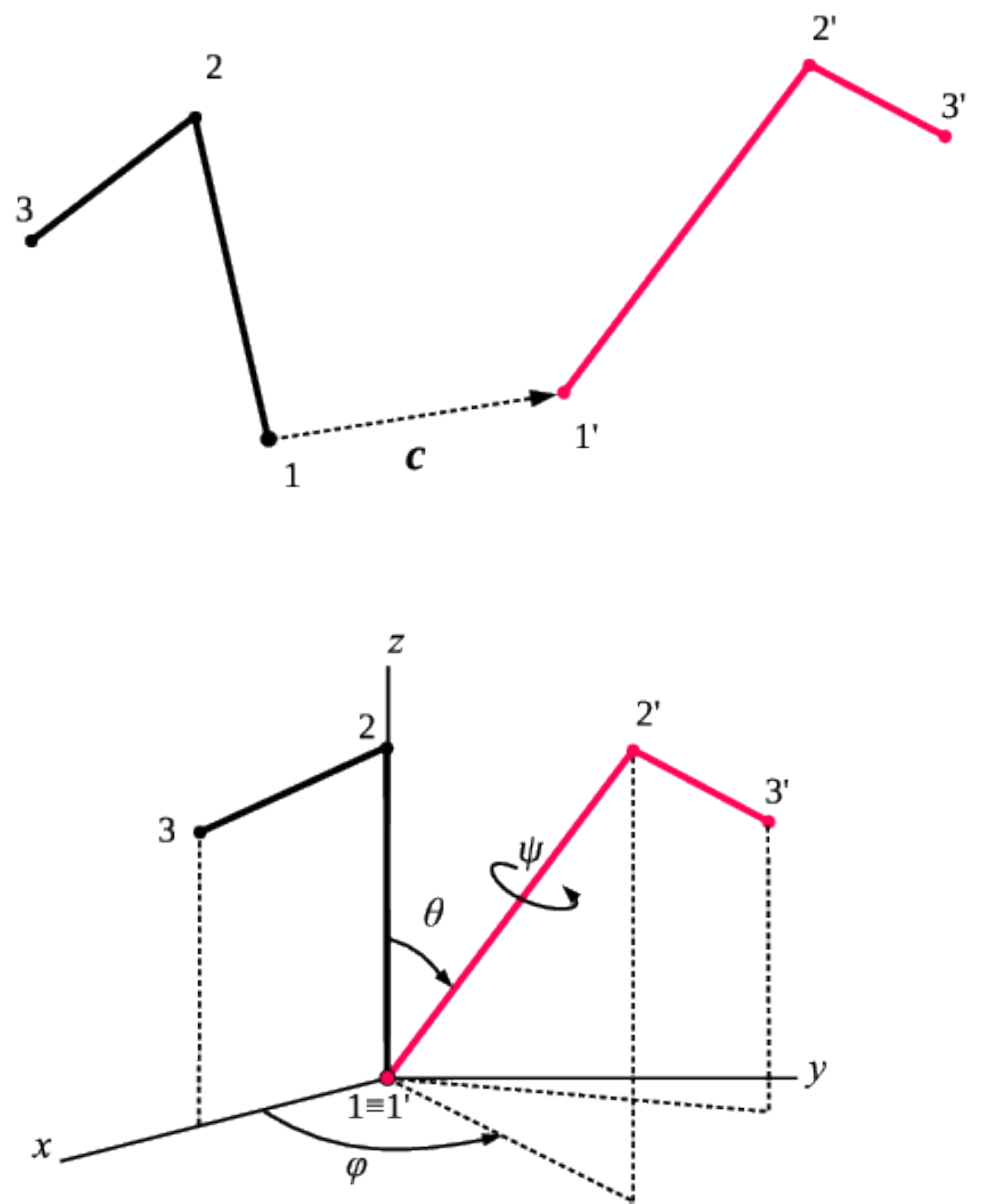}
\end{centering}
\caption{\label{fig:refframe} Diagrams illustrating the external degrees of freedom $\zeta = (\bm{c}, \bm{\omega})$, with $\bm{\omega} = (\theta, \varphi, \psi)$, of a target molecule (red) relative to a reference molecule (black). The reference frame of each molecule is defined by three non co-linear reference atoms denoted by $1, 2, 3$ and $1', 2', 3'$ for the reference and target molecules, respectively. $\mathbf{c}$ (top) is defined by the vector distance between $1'$ and $1$. The atoms $1, 2,$ and $3$ define a Cartesian coordinate system (bottom) with $1$ at the origin, and where the $z$ axis is along the $1$ to $2$ direction and the $x$ axis is oriented such that $3$ is in the $xz$ plane. When the target molecule is translated by $-\mathbf{c}$ so that $1'$ coincides to $1$, $\theta$ and $\varphi$ are the spherical polar angles of $2'$. The angle $\psi$ is defined by the dihedral angle formed by atoms $3$, $1$, $2'$, and $3'$ in this sequence. The alignment of the two molecules is maintained by restraining potentials such that $2'$ is approximately co-linear with $2$ ($\theta \simeq 0$)--in which case the $\varphi$ degree of freedom can be ignored--and $3'$ is approximately in the $xz$ plane ($\psi \simeq 0$). }
\end{figure}

We start with the statistical mechanics expression for the bimolecular binding constant between a receptor molecule $R$ and a ligand $A$\cite{Gilson:Given:Bush:McCammon:97,Gallicchio2011adv,Gallicchio2021binding}
\begin{equation}
K_{b}(A) =\frac{C^{\circ}}{8\pi^{2}}\frac{z_{RA}}{z_{R}z_{A}},
\label{eq:Kb_statmech}
\end{equation}
where
\begin{equation}
z_{A} = \int dx_A  e^{-\beta \Psi_{A}(x_{A})}
    \label{eq:part-func-zR}
\end{equation}
and similarly for R, and
\begin{equation}
z_{RA} = \int dx_R dx_A d\zeta_A I(\zeta_A) e^{-\beta \Psi_{RA}(x_{R}, x_{A}, \zeta_A)}
    \label{eq:part-func-zRL}
\end{equation}
where $x_R$ and $x_A$ are the internal coordinates of the receptor and ligand respectively, $\zeta_A = (\bm{c}_A, \bm{\omega}_A)$ denotes, collectively, the three position and three orientation coordinates of ligand $A$ relative, in this case, to the reference frame of the receptor (Fig.\ \ref{fig:refframe}), $\Psi_A$ and $\Psi_{RA}$ denote the effective potential energy functions in the solvent potential of mean force representation\cite{Roux:Simonson:99,Gallicchio2021binding} of the ligand in the solvent and of the ligand bound to the solvated receptor, respectively,  The function $I(\zeta_A)$ in Eq.\ (\ref{eq:part-func-zRL}) is an indicator function that is set to $1$ if the  position and orientation of the
ligand are such that receptor and ligand are considered bound, and zero otherwise.\cite{Gilson:Given:Bush:McCammon:97,Boresch:Karplus:2003,Gallicchio2011adv} The volume of configurational space encompassed by the indicator function is denoted here by $V_{\rm site} \Omega_{\rm site}$:
\begin{equation}
\int d\zeta I(\zeta) = V_{\rm site} \Omega_{\rm site}
\label{eq:vsite}
\end{equation}

Eq.\ (\ref{eq:Kb_statmech}) is turned into a statistical mechanics average by first considering the product $z_R z_A$ as the partition function of a system containing both $R$ and $A$, with $A$ located at a position $\zeta_A^\ast$ in the solvent bulk far away from $R$ so that it does not interact with it:
\begin{equation}
z_R z_{A} = \int dx_R dx_A  e^{-\beta \Psi_{RA}(x_{R},x_{A}, \zeta_A^\ast)}
    \label{eq:part-func-zR-zA-star}
\end{equation}
and then multiplying it and dividing it by
\begin{equation}
\int d\zeta_A^\ast I^\ast(\zeta_A^\ast) = V_{\rm site} \Omega_{\rm site}
    \label{eq:vsite-star}
\end{equation}
where
\begin{equation}
    I^\ast(\bm{c},\bm{\omega}) = I(\bm{c}-\bm{d},\bm{\omega})
    \label{eq:Istar-def}
\end{equation}
is an indicator function identical to the one that defines the complex, but centered at a point $\bm{d}$ in the solvent bulk relative to the reference frame of the receptor. 

After these manipulations, Eq.\ (\ref{eq:Kb_statmech}) becomes
\begin{equation}
    K_b(A) = C^\circ V_{\rm site} \frac{\Omega_{\rm site}}{8 \pi^2}
    \frac{
    \int dx_R dx_A d\bm{c}_A d\bm{\omega}_A I(\bm{c}_A , \bm{\omega}_A) e^{-\beta \Psi_{RA}(x_{R}, x_{A}, \bm{c}_A, \bm{\omega}_A)}
    }{
    \int dx_R dx_A d\bm{c}^\ast_A d\bm{\omega}_A I^\ast(\bm{c}^\ast_A , \bm{\omega}_A) e^{-\beta \Psi_{RA}(x_{R}, x_{A}, \bm{c}^\ast_A, \bm{\omega}_A)}
    }
    \label{eq:Kb-statmech-2}
\end{equation}
where the external coordinates $\zeta$ are expanded into their position and orientation components and, without loss of generality, the orientation dummy variables are denoted by $\bm{\omega}$ in both integrals. 
Next, we perform the change of variable $\bm{c}^\ast_A = \bm{c}_A + \bm{d}$, noting that, using Eq.\ (\ref{eq:Istar-def}),  $I^\ast(\bm{c}^\ast_A , \bm{\omega}_A) = I^\ast( \bm{c}_A + \bm{d}, \bm{\omega}_A) = I( \bm{c}_A, \bm{\omega}_A)$ to obtain
\begin{equation}
    K_b(A) = C^\circ V_{\rm site} \frac{\Omega_{\rm site}}{8 \pi^2}
    \frac{
    \int dx_R dx_A d\bm{c}_A d\bm{\omega}_A I(\bm{c}_A , \bm{\omega}_A) e^{-\beta \Psi_{RA}(x_{R}, x_{A}, \bm{c}_A, \bm{\omega}_A)}
    }{
    \int dx_R dx_A d\bm{c}_A d\bm{\omega}_A I(\bm{c}_A , \bm{\omega}_A) e^{-\beta \Psi_{RA}(x_{R}, x_{A}, \bm{c}_A + \bm{d}, \bm{\omega}_A)}
    }
    \label{eq:Kb-statmech-3}
\end{equation}
Finally, by multiplying and dividing the integrand in the numerator by \\ $\exp[-\beta \Psi_{RA}(x_{R}, x_{A}, \bm{c}_A + \bm{d}, \bm{\omega}_A)]$, Eq.\ (\ref{eq:Kb-statmech-3}) becomes
\begin{equation}
K_b(A) = C^\circ V_{\rm site} \frac{\Omega_{\rm site}}{8 \pi^2} \langle e^{-\beta u} \rangle_0
\label{eq:ATM-1}
\end{equation}
where $u$ is the perturbation function
\begin{equation}
u = \Psi_{RA}(x_{R}, x_{A}, \bm{c}_A, \bm{\omega}_A) - \Psi_{RA}(x_{R}, x_{A}, \bm{c}_A + \bm{d}, \bm{\omega}_A)
\label{eq:ATM-2}
\end{equation}
corresponding to the change in potential energy for rigidly translating the ligand from a position in the bulk $\bm{c}_A + \bm{d}$ to a position in the binding site $\bm{c}_A$, and the ensemble average $\langle \ldots \rangle_0$ is conducted when the ligand is in the bulk.

Eqs.\ (\ref{eq:ATM-1}) and (\ref{eq:ATM-2}) form the basis of the Alchemical Transfer Method (ATM) for the estimation of Absolute Binding Free Energies (ABFE).\cite{wu2021alchemical}

\subsection{Relative Binding Free Energy Formalism}
\label{sec:RBFEwithATM}

Consider now the ratio, $K_b({\rm B})/K_b({\rm A})$, of the equilibrium binding constants of two ligands A and B to the same receptor R in the same binding mode described by the indicator function $I(\zeta)$. To simplify the notation, in the following we will also assume that  $I(\zeta)$ is independent of the orientation coordinates. That is we assume that $I(\zeta) = I(\bm{c})$. Under this assumption $\Omega_{\rm site} = 8 \pi^2$ which cancels the same factor in the denominator of Eq.\ (\ref{eq:ATM-1}).

From Eq.\ (\ref{eq:Kb_statmech}) and canceling the common factor $z_{\rm R}$, we have
\begin{equation}
    \frac{K_b({\rm B})}{K_b({\rm A})} =
    \frac{z_{{\rm RB}} z_{\rm A}}{z_{{\rm RA}} z_{\rm B}} \, ,
    \label{eq:Kbratio1}
 \end{equation}
 where $z_{\rm RB}$ is the intramolecular configurational partition function of the complex between R and B, $z_{\rm B}$ is the intramolecular partition function of ligand B, and similarly for $z_{\rm RA}$ and $z_{\rm A}$. Analogously to manipulations that lead to Eq.\ (\ref{eq:Kb-statmech-2}), the products $z_{{\rm RB}} z_{\rm A}$ and  $z_{{\rm RA}} z_{\rm B}$ are expressed as partition function integrals of systems containing the receptor and the two ligands such that one of the ligands is in the binding site and the other is centered on a position of the bulk at distance $\bm{d}$ relative to the receptor coordinate frame. Specifically we multiply the numerator of Eq.\ (\ref{eq:Kbratio1}) by Eq.\ (\ref{eq:vsite-star}) for A and the denominator by the corresponding integral for B. The result is
  \begin{equation}
    \frac{K_b(B)}{K_b(A)} =
    \frac{
    \int dx_R dx_A dx_B d\bm{c}^\ast_A d\bm{\omega}_A d\bm{c}_B d\bm{\omega}_B I^\ast(\bm{c}^\ast_A) I(\bm{c}_B)  e^{-\beta \Psi(x_{R}, x_{A}, x_B, \bm{c}^\ast_A, \bm{\omega}_A, \bm{c}_B, \bm{\omega}_B)}
    }{
    \int dx_R dx_A dx_B d\bm{c}_A d\bm{\omega}_A d\bm{c}^\ast_B d\bm{\omega}_B I(\bm{c}_A) I^\ast(\bm{c}^\ast_B) e^{-\beta \Psi(x_{R}, x_{A}, x_B, \bm{c}_A, \bm{\omega}_A, \bm{c}^\ast_B, \bm{\omega}_B)}
    }
    \label{eq:Kb-statmech-rel-1}
\end{equation}
 Next, we perform the changes of variables $\bm{c}^\ast_A = \bm{c}_A + \bm{d}$ and  $\bm{c}_B = \bm{c}^\ast_B - \bm{d}$ for the integral in the numerator to obtain
 \begin{equation}
    \frac{K_b(B)}{K_b(A)} =
    \frac{
    \int dx_R dx_A dx_B d\bm{c}_A d\bm{\omega}_A d\bm{c}^\ast_B d\bm{\omega}_B I(\bm{c}_A) I^\ast(\bm{c}^\ast_B)  e^{-\beta \Psi(x_{R}, x_{A}, x_B, \bm{c}_A + \bm{d}, \bm{\omega}_A, \bm{c}^\ast_B - \bm{d}, \bm{\omega}_B)}
    }{
    \int dx_R dx_A dx_B d\bm{c}_A d\bm{\omega}_A d\bm{c}^\ast_B d\bm{\omega}_B I(\bm{c}_A) I^\ast(\bm{c}^\ast_B) e^{-\beta \Psi(x_{R}, x_{A}, x_B, \bm{c}_A, \bm{\omega}_A, \bm{c}^\ast_B, \bm{\omega}_B)}
    }
    \label{eq:Kb-statmech-rel-2}
\end{equation}
Again, by multiplying and dividing the integrand in the numerator by the Boltzmann factor in the denominator, Eq.\ (\ref{eq:Kb-statmech-rel-2}) is expressed by an ensemble average 
\begin{equation}
      \frac{K_b(B)}{K_b(A)} = \langle e^{-\beta u} \rangle_0
      \label{eq:Kbrel-average}
\end{equation}
with the perturbation energy 
\begin{equation}
u = \Psi(x_{R}, x_{A}, \bm{c}_A+\bm{d}, \bm{\omega}_A, \bm{c}^\ast_B-\bm{d}, \bm{\omega}_B ) - \Psi(x_{R}, x_{A}, \bm{c}_A, \bm{\omega}_A, \bm{c}^\ast_B, \bm{\omega}_B ) 
\label{eq:pert-func-rel}
\end{equation}
corresponding to the change in potential energy for rigidly translating ligand B from a position in the bulk $\bm{c}^\ast_B$ to a position in the binding site $\bm{c}_B - \bm{d}$ while at the same time ligand A is translated from its position in the binding site $\bm{c}_A$ to a position in the bulk $\bm{c}_A + \bm{d}$, and the ensemble average $\langle \ldots \rangle_0$ is conducted while ligand A is the binding site and ligand B is in the bulk.

Eqs.\ (\ref{eq:Kbrel-average}) and (\ref{eq:pert-func-rel}) are the basis of the Relative Binding Free Energy (RBFE) estimation protocol of the Alchemical Transfer Method (ATM) presented in this work. In analogy with the double-topology free energy perturbation framework,\cite{Mey2020Best} we call this approach \emph{dual-ligand} to emphasize the fact that the calculations is performed with a solvent box containing the two ligands.

\subsection{Alignment Restraints}
\label{sec:RBFEwithATMandRestraints}

While, Eq.\ (\ref{eq:Kbrel-average}) is formally correct, it converges slowly near the end-points because placing ligand B anywhere in the binding site and placing ligand A anywhere in the solvent bulk is likely to produce severe clashes with either receptor atoms or solvent molecules. For ligands of similar shape, the occurrences of severe clashes with nearby atoms would be reduced if the two ligands land in a similar location and orientation when their positions are exchanged. In order to improve convergence, we introduce a position restraint potential $U_r(c_{AB})$, as well as an orientation restraint potential $U_\Omega(\omega_{AB})$ to keep the two ligands at approximately the same distance and approximately aligned with each other.  Here, $c_{AB}$ represents the distance between the reference frames of the two ligands and $\omega_{AB}$ represents, schematically, their mutual orientation (Figure \ref{fig:refframe}).  
The specific form of the alignment restraint potentials used in this work is given below. For now, we simply assume that they are symmetric with respect to the exchange of the A and B ligand labels and we show that, under this assumption, the ATM RBFE formulation of Eqs.\ (\ref{eq:Kbrel-average}) and (\ref{eq:pert-func-rel}) is independent of the specific alignment restraint potential. 

To see why this is the case, note that to obtain Eq.\ (\ref{eq:Kb-statmech-rel-1}), we multiplied and divided Eq.\ (\ref{eq:Kbratio1}) by Eq.\ (\ref{eq:vsite-star}). By instead multiplying the numerator by the integrals
\begin{equation}
    V_r = \int d\bm{c}^\ast_A I^\ast(\bm{c}^\ast_A) e^{-\beta U_r(c_{AB})}
    \label{eq:VrA}
\end{equation}
and
\begin{equation}
    \Omega_r = \int d\bm{\omega}_A e^{-\beta U_\Omega(\omega_{AB})}
\label{eq:OmegarA}
\end{equation}
and by multiplying the denominator by the corresponding integrals for ligand B that have the same values as those for ligand A, we obtain (notice that the integrals $V_r$ and $\Omega_r$ above and the corresponding ones for ligand B, are constants independent of the external degrees of freedom of the ligand that is not integrated over) 
  \begin{equation}
    \frac{K_b(B)}{K_b(A)} =
    \frac{
    \int dx_R dx_A dx_B d\bm{c}^\ast_A d\bm{\omega}_A d\bm{c}_B d\bm{\omega}_B  I^\ast(\bm{c}^\ast_A) I(\bm{c}_B)
    e^{-\beta [ U_r(c_{AB}) +  U_\Omega(\omega_{AB}) ] }
     e^{-\beta \Psi(x_{R}, x_{A}, x_B, \bm{c}^\ast_A, \bm{\omega}_A, \bm{c}_B, \bm{\omega}_B)}
    }{
    \int dx_R dx_A dx_B d\bm{c}_A d\bm{\omega}_A d\bm{c}^\ast_B d\bm{\omega}_B 
    I(\bm{c}_A) I^\ast(\bm{c}^\ast_B)
     e^{-\beta [ U_r(c_{AB}) +  U_\Omega(\omega_{AB} ) ] }
    e^{-\beta \Psi(x_{R}, x_{A}, x_B, \bm{c}_A, \bm{\omega}_A, \bm{c}^\ast_B, \bm{\omega}_B)}
    }
    \label{eq:Kb-statmech-rel-restr-1}
\end{equation}
As before, we set $\bm{c}^\ast_A = \bm{c}_A + \bm{d}$ and  $\bm{c}_B = \bm{c}^\ast_B - \bm{d}$ in the numerator, then use Eq.\ (\ref{eq:Istar-def}) and the symmetry property of the restraint potential functions to obtain:
 \begin{equation}
    \frac{K_b(B)}{K_b(A)} =
    \frac{
    \int dx_R dx_A dx_B d\bm{c}_A d\bm{\omega}_A d\bm{c}^\ast_B d\bm{\omega}_B 
     I(\bm{c}_A) I^\ast(\bm{c}^\ast_B)
    e^{-\beta [ U_r(c_{AB}) +  U_\Omega(\omega_{AB}) ] }
     e^{-\beta \Psi(x_{R}, x_{A}, x_B, \bm{c}_A + \bm{d}, \bm{\omega}_A, \bm{c}^\ast_B - \bm{d}, \bm{\omega}_B)}
    }{
    \int dx_R dx_A dx_B d\bm{c}_A d\bm{\omega}_A d\bm{c}^\ast_B d\bm{\omega}_B 
    I(\bm{c}_A)   I(\bm{c}^\ast_B)
     e^{-\beta [ U_r(c_{AB}) +  U_\Omega(\omega_{AB}) ] }
    e^{-\beta \Psi(x_{R}, x_{A}, x_B, \bm{c}_A, \bm{\omega}_A, \bm{c}^\ast_B, \bm{\omega}_B)}
    }
    \label{eq:Kb-statmech-rel-restr-2}
\end{equation}
which is the same as Eq.\ (\ref{eq:Kb-statmech-rel-2}) with the addition of the alignment restraint potential terms. The corresponding ensemble average expression is the same as Eq.\ (\ref{eq:Kbrel-average}) with Eq.\ (\ref{eq:pert-func-rel}), except that A and B remain aligned. Eq.\ (\ref{eq:Kb-statmech-rel-restr-2}) confirms that the ATM RBFE formula Eq.\ (\ref{eq:Kbrel-average}) yields an unbiased estimate of the relative binding free energy with or without alignment restraints. 

The specific forms of the restraining potentials used in this work are as follows. The positional restraint is a spherical harmonic potential between the reference atoms $1$ and $1'$ of the two ligands:
\begin{equation}
    U_r(c_{AB}) = \frac12 k_r |\bm{r}^\ast_{1A} - \bm{d} - \bm{r}_{1B} |^2
    \label{eq:pos-restraint}
\end{equation}
where $\bm{r}^\ast_{1A}$ is the position of the first reference atom of ligand A in the bulk and $\bm{r}_{1B}$ is the position of the first reference atom of ligand B in the binding site. Given the relationship $\bm{r}^\ast = \bm{r} + \bm{d}$, this potential is symmetric with respect to the exchange of A and B. The orientation restraints are implemented by two  potentials. One, named $U_\theta$, is based on the angle $\theta$ between the $z$-axis of the reference frame of A relative to that of B (Fig.\ \ref{fig:refframe}), which is naturally symmetric with respect to the exchange of the two frames. The second restraint is based on the dihedral angle $\psi$ in Figure \ref{fig:refframe}. Because the $\psi$ angle is not symmetric with respect to the exchange of A and B, we use a symmetrized restraining potential of the form
\begin{equation}
    \frac12 [ U_\psi(\psi_{BA}) +  U_\psi(\psi_{AB}) ]
\end{equation}
where $U_\psi$ is the restraining potential below, $\psi_{BA}$ and $\psi_{AB}$ are the $\psi$ angles of the reference frame of ligand B with respect to A and of A relative to B, respectively. Each orientation restraint potential is based on the cosine of the angle. For example
\begin{equation}
     U_\theta(\theta) = \frac12 k_\theta (1 - \cos\theta)
     \label{eq:angle-restraint}
\end{equation}
and similarly for the $\psi$ angle, where $k_\theta$ (and $k_\psi$ in the case of the $\psi$ angle) is a dimensionless factor that controls the strength of the restraint.

\subsection{Calculation Protocol}

For convenience of notation, Eqs.\ (\ref{eq:Kbrel-average}) and (\ref{eq:pert-func-rel}), were derived in the solvent potential of mean force formalism to avoid carrying the degrees of freedom of the solvent. When the definition of the solvent potential of mean force is substituted into Eq.\ (\ref{eq:Kb-statmech-rel-2}) to show explicitly the degrees of freedom of the solvent,\cite{Gallicchio2011adv} the RBFE formula in Eq.\ (\ref{eq:Kbrel-average}) is unchanged and the formula for the perturbation  energy Eq.\ (\ref{eq:pert-func-rel}) becomes 
\begin{equation}
    u = U(x_{R}, x_{A}, \bm{c}_A+\bm{d}, \bm{\omega}_A, \bm{c}^\ast_B-\bm{d}, \bm{\omega}_B, x_S ) - U(x_{R}, x_{A}, \bm{c}_A, \bm{\omega}_A, \bm{c}^\ast_B, \bm{\omega}_B, x_S ) 
\label{eq:pert-func-rel-expl}
\end{equation}
in which the actual potential energy function $U$ replaces the effective potential energy function $\Psi$, and, in addition to the coordinates of the receptor and the ligands, it includes the coordinates, $x_S$, of the solvent.

While Eqs.\ (\ref{eq:Kbrel-average}) and (\ref{eq:pert-func-rel}) formally yield the correct relative binding free energy, their direct implementation in terms of the exponential average of the perturbation energy is numerically unstable.\cite{Gallicchio2021binding} Instead, the free energy change is computed by well-established protocols based on stratification using a sequence of alchemical states defined by a $\lambda$-dependent Hamiltonian\cite{Chipot:Pohorille:book:2007} and multi-state thermodynamic reweighting methods.\cite{Shirts2008a,Tan2012} The specific alchemical protocol we employ is described in detail in previous publications.\cite{pal2019perturbation,khuttan2021single,wu2021alchemical} Briefly, in this work we employ a linear $\lambda$-dependent alchemical potential energy function of the form
\begin{equation}
U_{\lambda}(x)=U_{0}(x)+W_{\lambda}(u)\label{eq:pert_pot}
\end{equation}
where $x$ represents the set of atomic coordinates of the system (the internal coordinates of the ligands and the receptor, the external coordinates of the ligands, and the coordinates of the solvent), the perturbation energy $u(x)$ is defined by Eq.\ (\ref{eq:pert-func-rel-expl})
and $W_{\lambda}(u)$ is the generalized softplus alchemical perturbation function
\begin{equation}
  W_{\lambda}(u)=\frac{\lambda_{2}-\lambda_{1}}{\alpha}\ln\left[1+e^{-\alpha(u_{\rm sc}(u)-u_{0})}\right]+\lambda_{2}u_{\rm sc}(u)+w_{0} .
  \label{eq:ilog-function}
\end{equation}
The parameters $\lambda_{2}$, $\lambda_{1}$, $\alpha$, $u_{0}$, and $w_{0}$ are functions of $\lambda$ which are adjusted to enhance convergence,\cite{khuttan2021single} and 
\begin{equation}
  u_{\rm sc}(u)=
\begin{cases}
u & u \le u_c \\
(u_{\rm max} - u_c ) f_{\rm sc}\left[\frac{u-u_c}{u_{\rm max}-u_c}\right] + u_c & u > u_c
\end{cases}
\label{eq:soft-core-general}
\end{equation}
with
\begin{equation}
f_\text{sc}(y) = \frac{z(y)^{a}-1}{z(y)^{a}+1} \label{eq:rat-sc} \, ,
\end{equation}
and
\begin{equation}
    z(y)=1+2 y/a + 2 (y/a)^2
\end{equation}
is the soft-core perturbation energy function to avoid singularities near the initial state of the alchemical transformation.\cite{pal2019perturbation} The parameters $u_{\rm max}$, $u_c$, and $a$ are set to cap the perturbation energy $u(x)$ to a maximum positive value without affecting it away from the singularity. The specific values of $u_0$, $u_{\rm max}$, and of the scaling parameter $a$ used in this work are listed in the Computational Details.

The standard linear alchemical interpolation potential
\begin{equation}
U_{\lambda}(x)=U_{0}(x)+ \lambda u_{sc}[u(x)]
\label{eq:pert_pot_linear}
\end{equation}
is a special case of Eq.\ (\ref{eq:ilog-function}) for $\lambda_1 = \lambda_2 = \lambda$. With one exception (see Calculation Details) all of the RBFE calculations reported in this work employed the linear alchemical potential. 

Similar to the ABFE protocol,\cite{wu2021alchemical} the RBFE alchemical protocol involves two free energy legs. The initial state of the first leg corresponds to the physical state in which ligand 1 is bound to the receptor and ligand 2 is in the solvent bulk. The initial state of the second leg is the physical state in which the positions of the two ligands are swapped relative to the first leg. The initial states of the two legs are connected to the same alchemical intermediate defined by the potential function (\ref{eq:pert_pot}) at $\lambda = 1/2$ corresponding to a non-physical state in which both ligands are present simultaneously in the binding site and in the solvent bulk, while not interacting with each other  but each only interacting at 50\% strength with the respective environments. The difference of the free energy changes, $\Delta G_1$ and $\Delta G_2$, of each leg gives the relative standard binding free energy of the two ligands
\begin{equation}
    \Delta \Delta G^\circ_b(1,2) = \Delta G^\circ_b(2) - \Delta G^\circ_b(1) = \Delta G_1 - \Delta G_2 
\end{equation}

The RBFE alchemical calculation for leg 1, for example, consists of a series of molecular dynamics simulations with the potential function (\ref{eq:pert_pot}) distributed at increasing values of $\lambda$ varying from $0$ to $1/2$ corresponding to ligand 1 bound and ligand 2 unbound, and to the alchemical intermediate, respectively. At each MD step, the perturbation energy (\ref{eq:pert-func-rel-expl}) is obtained by displacing ligand 1 into the solvent bulk by translating its position along the vector $h$ while simultaneously translating ligand 2 in the opposite direction. The values of the perturbation energies are saved at regular intervals and processed by conventional multi-state reweighting analysis\cite{Tan2012} to compute the free energy change. The simulations for leg 2 are performed in the same way but starting from placing ligand 2 is in the binding site and ligand 1 is in the solvent bulk.

\subsection{Molecular Systems}

We have tested the ATM RBFE protocol on the SAMPL8 GDCC host-guest set\cite{Azimi2021SAMPL8}\footnote{https://github.com/\-samplchallenges/\-SAMPL8/\-tree/\-master/\-host\_guest/\-GDCC}  and on a set of complexes of the estrogen receptor $\alpha$ (ER$\alpha$).\cite{norman2006benzopyrans,wang2017accurate}

\begin{figure}[bt]
    \centering
    \includegraphics[scale=0.75]{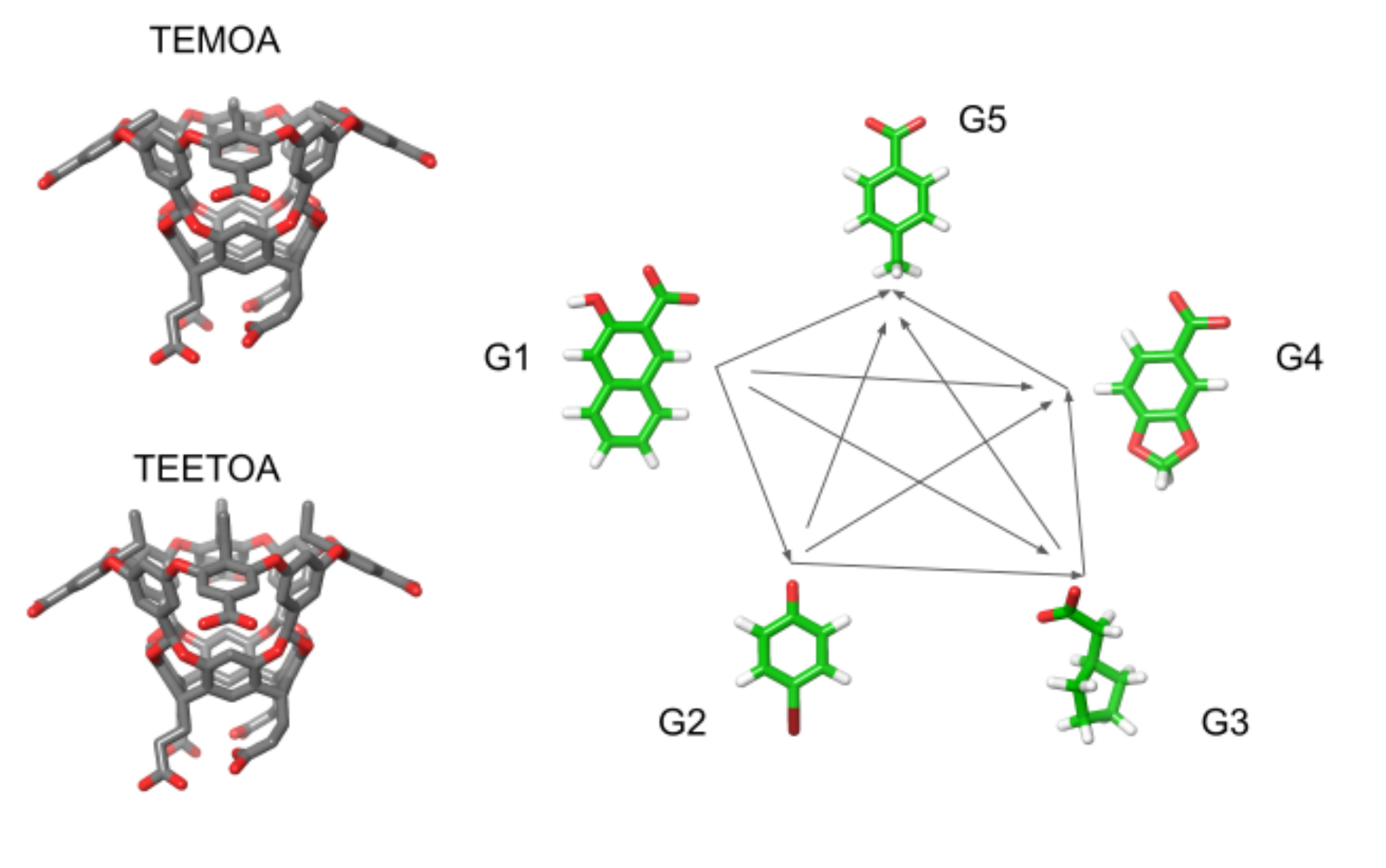}
    \caption{The SAMPL8 GDCC molecular systems}
    \label{fig:SAMPL8-systems}
\end{figure}

The SAMPL8 GDCC host-guest set consists of five small molecule guests binding to two host receptors (Figure \ref{fig:SAMPL8-systems}). The hosts and the guests are known to form complexes with 1:1 stoichiometry under experimental conditions. The GDCC hosts (named TEMOA and TEETOA) are cup-shaped with a deep hydrophobic cavity occupied by the non-polar end of the bound guests. In the host-guest complexes, the charged/polar groups of the guests tend to be oriented towards the solvent. The hosts differ in the nature of the sidechains that surround the rim of the cavity (methyl groups in TEMOA and longer ethyl groups in TEETOA). In this work, all pairwise relative binding free energies for this set were compared to the differences of the ATM absolute binding free energy estimates that were obtained previously.\cite{Azimi2021SAMPL8} The ABFE estimates for this set are considered as a particularly trustworthy reference because they closely agree with independent potential of mean force estimates and with the experimentally measured binding free energies.\cite{Azimi2021SAMPL8} As shown in Figure \ref{fig:SAMPL8-systems}, the majority of the pairwise RBFE calculations in this set involve scaffold hopping transformations. For example, all of the transformations to or from the G3 guest, among others, involve the deletion of a ring or the conversion of an aromatic 6-membered ring to an aliphatic 5-memberd ring. All of the guests except G2 are modeled in their anionic deprotonated forms. The G2 guest is modeled as neutral with the phenol oxygen protonated, as in our previous work, the protonated form of G2 demonstrated to contribute more to binding to both TEMOA and TEETOA hosts.\cite{Azimi2021SAMPL8} Hence, because all of the transformations involving G2 involve a change of the net charge of the ligand, this set tests the ability of the ATM RBFE protocol to handle changes in net charge in addition to changes of ligand topology. 

\begin{figure}[bt]
    \centering
    \includegraphics[scale=0.75]{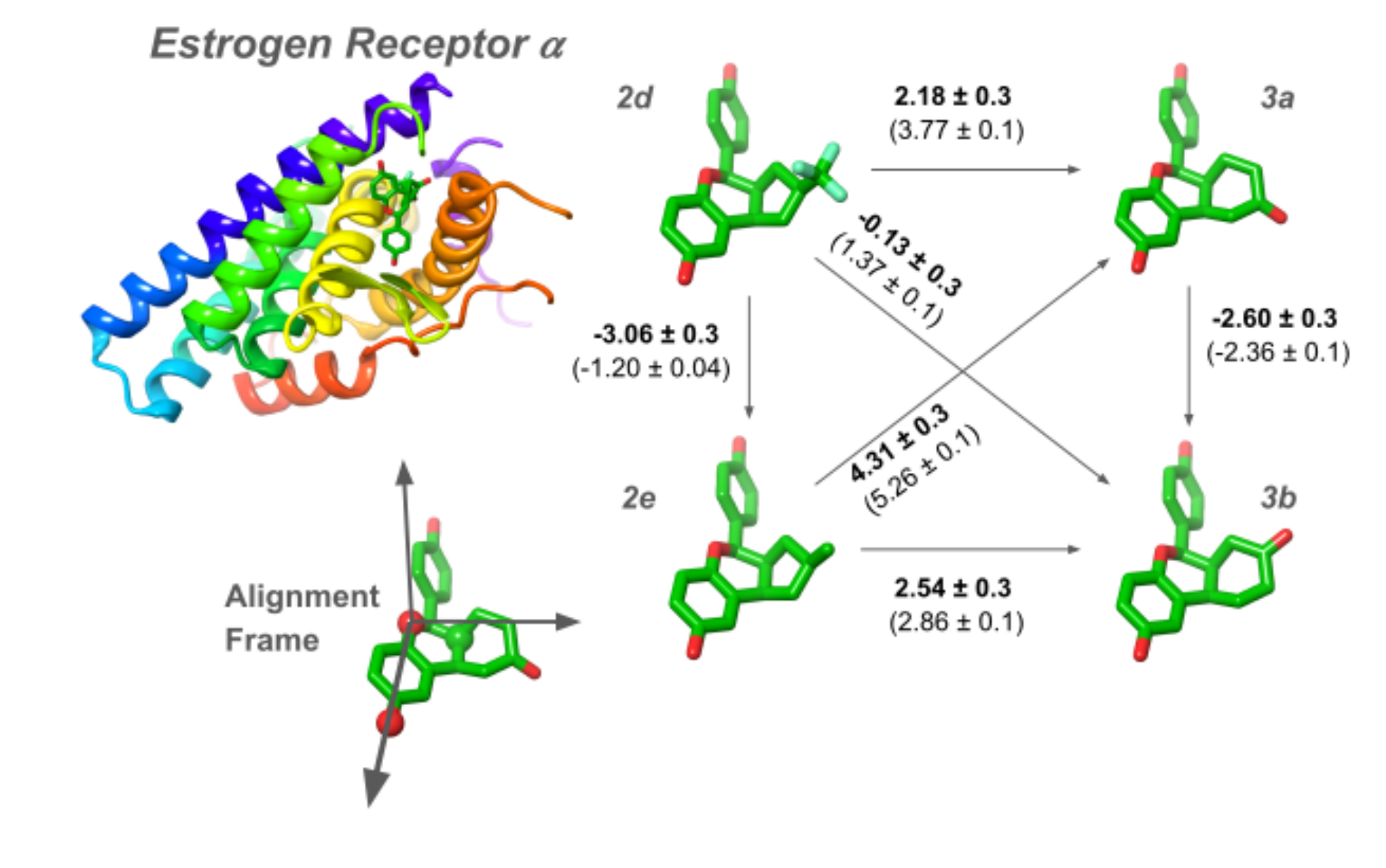}
    \caption{The ERalpha complexes}
    \label{fig:ERalpha-systems}
\end{figure}

The ER$\alpha$ benchmark consists of four molecular complexes of the ER$\alpha$ receptor with a series of benzopyran agonists.\cite{norman2006benzopyrans} The four ligands (named 2e, 2d, 3a, and 3b) bind in a similar fashion to a large solvent-occluded cavity of the receptor (Figure \ref{fig:ERalpha-systems}). The ligands share a similar topology consisting of a common scaffold along the long molecular axis ending in two polar phenol groups, flanked by a variable side group with either a five- or six-membered ring. Alchemical RBFE transformations in this set can be cleanly classified as either straightforward terminal R-group modifications (such as 3a to 3b) or more challenging scaffold-hopping transformations involving ring expansion (such as 2e to 3a). The common long molecular axis and the conserved portion of the side group provide a natural definition of the reference frame to the four ligands.\ref{fig:ERalpha-systems}

\subsection{Computational Details}

The host-guest systems were prepared from the provided MOL2 files provided by the SAMPL8 organizers ({\tt https://github.com/\-samplchallenges/\-SAMPL8/\-tree/\-master/\-host\_guest/\-GDCC}). The guests were manually docked to the hosts and aligned relative to each other using  Maestro (Schr\"{o}dinger Inc.). The structures of the ER$\alpha$ protein complexes were also prepared in Maestro starting from the provided files.\cite{wang2017accurate} Force-field parameter assignments and solvation of the molecular systems were performed using Ambertools 18. The GAFF1.8/AM1-BCC force field parameters\cite{wang2006automatic,he2020fast} were assigned for the hosts and the ligands and the Amber ff14SB parameters\cite{maier2015ff14sb,zou2019blinded} were assigned to the ER$\alpha$ protein receptor. The complexes were assembled using the tleap program in Ambertools 18. Prior to adding the solvent, the receptor and a pair of aligned ligands were loaded into tleap where one of the ligands was translated to a position corresponding to the solvent bulk (see below). Solvent molecules were added with a 10 \AA\ buffer around the complex and the unbound ligand of the ligand pair. The resulting solvent box was neutralized by adding sodium or chloride ions as needed. 

The complexes were energy minimized and thermalized at 300 K. Beginning at the bound state at $\lambda=0$, the systems were then annealed to the symmetric alchemical intermediate at $\lambda = 1/2$ for $250$ ps using the ATM alchemical potential energy function for Leg 1 [Eq.\ (\ref{eq:pert_pot_linear})]. This step provides a suitable initial configuration of the system without severe unfavorable repulsion interactions at alchemical intermediate state to seed the molecular dynamics simulations. The host molecules were loosely restrained with a flat-bottom harmonic potential of force constant 25.0 kcal/(mol \AA$^2$) and a tolerance of 1.5 \AA\ was set on the heavy atoms at the lower cup of the molecule (the first 40 atoms of the host as listed in the provided files). The same positional restraints were applied to the C$\alpha$ atoms of the ER$\alpha$ protein receptor.

With one exception (see below), the linear alchemical perturbation potential, $W_\lambda(u) = \lambda u_{\rm sc}(u)$ was used for all alchemical calculations with $11$ $\lambda$-states uniformly distributed between $\lambda=0$ and $1/2$ for each of the two ATM legs. The calculation for the more challenging test without alignment restraints from TEMOA-G4 to TEMOA-G1  employed the softplus perturbation potential (\ref{eq:pert_pot}) with the parameters listed in Table \ref{tab:G1-G4}.

\begin{table}[tb]
\begin{centering}
  \caption{\label{tab:G1-G4} Alchemical schedule of the softplus perturbation function for the two legs of the TEMOA-G1 to TEMOA-G4 transformation.}
\begin{tabular}{lccccc}
  $\lambda$  &   $\lambda_1$ & $\lambda_2$ & $\alpha$$^a$ & $u_0$$^b$ & $w_0$$^b$  \tabularnewline \hline
0.00 & 0.00 & 0.00 & 0.10 & 150 & 0 \tabularnewline
0.05 & 0.00 & 0.05 & 0.10 & 135 & 0 \tabularnewline
0.10 & 0.00 & 0.10 & 0.10 & 120 & 0 \tabularnewline
0.15 & 0.00 & 0.15 & 0.10 & 105 & 0 \tabularnewline
0.20 & 0.00 & 0.20 & 0.10 &  90 & 0 \tabularnewline
0.25 & 0.00 & 0.25 & 0.10 &  75 & 0 \tabularnewline
0.30 & 0.10 & 0.30 & 0.10 &  60 & 0 \tabularnewline
0.35 & 0.20 & 0.35 & 0.10 &  40 & 0 \tabularnewline
0.40 & 0.30 & 0.40 & 0.10 &  40 & 0 \tabularnewline
0.45 & 0.40 & 0.45 & 0.10 &  40 & 0 \tabularnewline
0.50 & 0.50 & 0.50 & 0.10 &  40 & 0 \tabularnewline
\hline
\end{tabular} 
\end{centering}
\begin{flushleft}\small
$^a$In (kcal/mol)$^{-1}$. $^b$In kcal/mol.
\end{flushleft}
\end{table}

The soft-core perturbation energy Eq.\ (\ref{eq:soft-core-general}) was used for all calculations with $u_{\rm max}=200 $ kcal/mol, $u_c=100 $ kcal/mol, and $a=1/16$.  The ligands was sequestered within the binding site by means of a flat-bottom harmonic potential between the centers of mass of the receptor and the ligand with a force constant of $25$ kcal/mol {\AA}$^2$ applied for separation greater than $4.5$ {\AA}. Perturbation energy samples and trajectory frames were saved every 10 ps. Asynchronous Hamiltonian replica exchanges\cite{gallicchio2015asynchronous} in $\lambda$-space were performed every 10 ps. The Langevin thermostat with a time constant of 2 ps was used to maintain the temperature at 300 K. Each replica was simulated for a minimum of 20 ns for the host-guest systems and for 14 ns for the protein-ligand systems. Binding free energies and the corresponding uncertainties were computed from the perturbation energy samples using UWHAM\cite{Tan2012}, discarding the first half of the trajectory. 

Alignment restraints were imposed using three reference atoms of the ligands. For the SAMPL8 guests, alignment settings were customized for each pair to maintain an approximate alignment between the the polar groups oriented towards the solvent and the body of the guest oriented towards the binding cavity of the host. The values of the force constants for the restraining potentials in Eqs.\ (\ref{eq:pos-restraint}) and (\ref{eq:angle-restraint}) were adjusted to allow for more flexibility between more dissimilar guests. $k_r$ varied from $25$ to $0.5$ kcal/mol/\AA$^2$, $k_\theta$ varied from  $50$ to $ 10$ kcal/mol, and $k_\psi$ varied  from $50$ to $ 0$ kcal/mol. The reference atoms for the ER$\alpha$ ligands are illustrated in Figure \ref{fig:ERalpha-systems} and the corresponding force constants were set to $k_r = 2.5$ kcal/mol/\AA$^2$, $k_\theta = 10$ kcal/mol, and $k_\psi = 10$ kcal/mol. The specific alignment settings for each system can be found in the simulation input files posted at {\tt github.com/\-Gallicchio-Lab/\-ATM-relative-binding-free-energy-paper}.

The alchemical molecular dynamics simulations were performed with the OpenMM 7.3\cite{eastman2017openmm} MD engine and the SDM integrator plugin ({\tt github.com/\-Gallicchio-Lab/\-openmm\_sdm\_plugin.git}) using the OpenCL platform. The ASyncRE software,\cite{gallicchio2015asynchronous} customized for OpenMM and SDM ({\tt github.com/\-Gallicchio-Lab/\-async\_re-openmm.git}), was used for the Hamiltonian Replica Exchange in $\lambda$ space for each ATM leg. Simulation input files for this work are freely available at {\tt github.com/\-Gallicchio-Lab/\-ATM-relative-binding-free-energy-paper}. The replica exchange simulations were performed on the XSEDE Comet and Expanse GPU HPC cluster at the San Diego Supercomputing Center, each using four GPUs per node.

\section{Results}

\subsection{SAMPL8 Host-Guest Systems}

\begin{figure}
    \centering
    \includegraphics[scale=0.8]{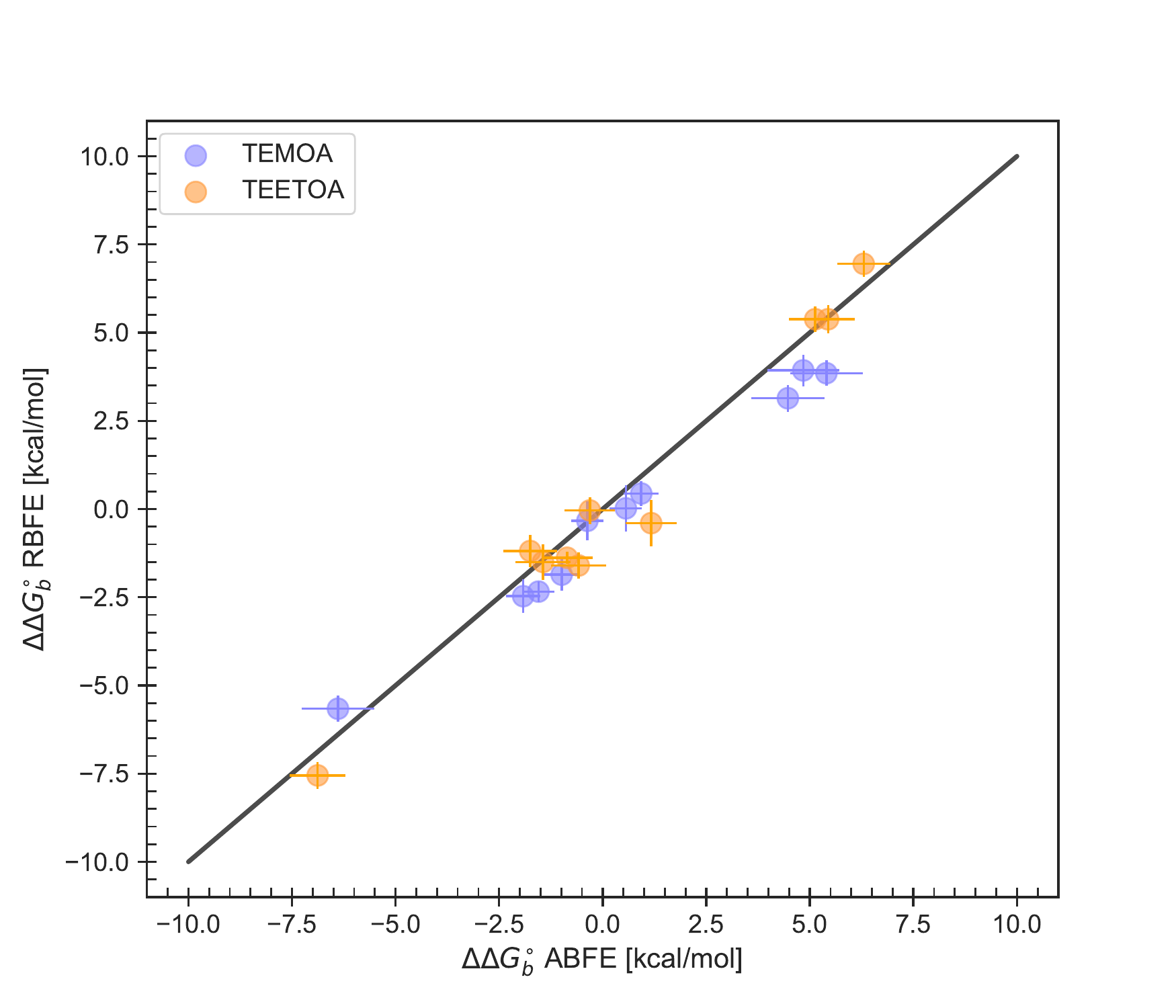}
    \caption{Validation of RBFE against the ABFE values for the SAMPL8 benchmark set. The line represents perfect agreement between the RBFE values and the differences of the corresponding ABFE results with the root mean square deviation of 0.8 kcal/mol, a value well within statistical uncertainties . 
}
    \label{fig:reg-SAMPL8}
\end{figure}

Table \ref{tab:rbfe-sampl8} and Figure \ref{fig:reg-SAMPL8} report the calculated relative binding free energies for the SAMPL8 host-guest systems compared to the corresponding estimates from the absolute binding free energies reported in reference \citenum{Azimi2021SAMPL8} and reproduced here in Table \ref{tab:ATM-simulation-results}. The results of the tests of the alignment restraints on the TEMOA-G1-G4 system are shown in Table \ref{tab:g1-g4-restraints}.

As shown in Table \ref{tab:rbfe-sampl8}, the calculated RBFEs for the twenty pairs of complexes (5th column), ten for each SAMPL8 host, are in good agreement with the corresponding estimates from the ABFEs values (4th column). The root mean square deviation between the two sets of free energy differences is only 0.8 kcal/mol, a value well within statistical uncertainties. The statistical uncertainties of the RBFEs estimates are found to be systematically smaller than those of the differences of ABFEs estimates. For example, the uncertainty for the large G1 to G3 scaffold-hopping transformation is 0.38 kcal/mol compared to an uncertainty of 0.66 kcal/mol from the differences of the corresponding ABFE estimates. The RBFE estimates are also found to have a high level of self-consistency with an average 3-points cycle closure error of only 0.25 kcal/mol. The latter value is computed by evaluating the free energy change along all of the triangular cycles in the network of transformations in Figure \ref{fig:SAMPL8-systems}.

As illustrated in Table \ref{tab:rbfe-sampl8} the RBFE estimates are obtained from the differences between the free energy changes for the two legs ($\Delta G_1$ and $\Delta G_2$ in Table \ref{tab:rbfe-sampl8}). These tend to be of moderate magnitude for the transformations between guests with the same net charge--between 10 to 20 kcal/mol range in magnitude, compared to the much larger free energy changes for the two legs of the ABFE calculations, which range between 40 to 50 kcal/mol.\cite{Azimi2021SAMPL8} This trend, which is remarkably maintained even for large scaffold-hopping transformations such as G1 to G3,  can be justified by the large desolvation penalty occurring for binding a negatively-charged guest to the host that cancels out in the RBFE calculations. Conversely, the trend is reversed for the RBFE transformations involving the neutral protonated G2 guest (G2P in Tables \ref{tab:rbfe-sampl8} and \ref{tab:ATM-simulation-results}), in which the ABFE leg free energies tend to be significantly smaller than for the RBFE transformations. Evidently, in these cases, the desolvation penalty is much smaller whereas the RBFE transformations are contending with a change of net charge. Nevertheless, even in these cases the RBFE calculations yield consistent free energy estimates with reasonably small statistical uncertainties.

\begin{table}[tb]
\centering
\caption{ATM absolute binding free energy estimates for the host-guest complexes from reference \citenum{Azimi2021SAMPL8}.}
\label{tab:ATM-simulation-results}
\begin{tabular}{lc}
Complex    & $\Delta G^\circ_b$  \\ \hline 
TEMOA-G1   & $ -6.71  \pm 0.30 $ \\ 
TEMOA-G2P  & $-13.10  \pm 0.83 $ \\
TEMOA-G3   & $ -8.26  \pm 0.25 $ \\ 
TEMOA-G4   & $ -8.63  \pm 0.30 $ \\ 
TEMOA-G5   & $ -7.70  \pm 0.30 $ \\ \hline
TEETOA-G1  & $ -1.07  \pm 0.34 $ \\ 
TEETOA-G2P & $ -7.95  \pm 0.28 $ \\ 
TEETOA-G3  & $ -1.65  \pm 0.30 $ \\ 
TEETOA-G4  & $ -2.51  \pm 0.30 $ \\ 
TEETOA-G5  & $ -2.82  \pm 0.28 $ \\ \hline
\end{tabular}
\end{table}

\begin{table}[tb]
    \centering
    \caption{Relative binding free energies of the TEMOA and TEETOA host-guest complexes.}
    \label{tab:rbfe-sampl8}
    \begin{tabular}{lcccc}
Complex Pair & $\Delta G_1$$^{a,b}$      & $\Delta G_2$$^{a,b}$     & $\Delta\Delta G^\circ_{\rm b}$(RBFE)$^{a,b}$ & $\Delta\Delta G^\circ_b$(ABFE)$^{a,c}$ \\ \hline
 TEMOA-G1-G2p & $41.68 \pm 0.21$  & $47.34 \pm 0.30$ & $ -5.66 \pm 0.37$ & $ -6.39 \pm 0.88$\\ 
 TEMOA-G1-G3  & $20.67 \pm 0.18$  & $23.01 \pm 0.24$ & $ -2.34 \pm 0.30$ & $ -1.55 \pm 0.39$\\ 
 TEMOA-G1-G4  & $15.98 \pm 0.30$  & $18.45 \pm 0.39$  & $ -2.47 \pm 0.49$ & $ -1.92 \pm 0.42$\\ 
 TEMOA-G1-G5  & $14.53 \pm 0.24$  & $16.39 \pm 0.39$ & $ -1.86 \pm 0.46$ & $ -0.99 \pm 0.42$\\ 
 TEMOA-G2p-G3 & $49.52 \pm 0.30$  & $45.59 \pm 0.36$  & $  3.93 \pm 0.47$ & $ 4.84 \pm 0.87$\\ 
 TEMOA-G2p-G4 & $45.19 \pm 0.27$  & $42.05 \pm 0.27$ & $3.14 \pm 0.38$ & $  4.47 \pm 0.88$\\ 
 TEMOA-G2p-G5 & $44.07 \pm 0.24$  & $40.22 \pm 0.27$ & $  3.85 \pm 0.36$ & $ 5.40 \pm 0.88$\\ 
 TEMOA-G3-G4  & $16.62 \pm 0.39$  & $16.95 \pm 0.39$ & $ -0.33 \pm 0.55$ & $ -0.37 \pm 0.39$ \\ 
 TEMOA-G3-G5  & $16.82 \pm 0.48$  & $16.80 \pm 0.45$ & $ 0.02 \pm 0.66$ & $0.56 \pm 0.39$\\ 
 TEMOA-G4-G5  & $10.80 \pm 0.27$  & $10.35 \pm 0.24$ & $ 0.44 \pm 0.36$ & $0.93 \pm 0.42$ \\ 
 
TEETOA-G1-G2p & $40.97 \pm 0.27$  & $48.52 \pm 0.27$ & $ -7.55 \pm 0.38$ &$ -6.88 \pm 0.67$\\ 
TEETOA-G1-G3  & $21.64 \pm 0.24$  & $23.24 \pm 0.27$ & $ -1.60 \pm 0.36$ & $ -0.58 \pm 0.66$ \\ 
TEETOA-G1-G4  & $16.20 \pm 0.30$  & $17.70 \pm 0.39$ & $ -1.50 \pm 0.49$ & $ -1.44 \pm 0.66$ \\
TEETOA-G1-G5  & $14.51 \pm 0.24$  & $15.70 \pm 0.39$ & $ -1.19 \pm 0.46$ & $ -1.75 \pm 0.65$\\ 
TEETOA-G2p-G3 & $51.89 \pm 0.27$  & $44.94 \pm 0.27$ & $ 6.95 \pm 0.38$ & $ 6.30 \pm 0.61$\\ 
TEETOA-G2p-G4 & $46.33 \pm 0.30$ & $40.95 \pm 0.27$ & $ 5.38 \pm 0.40$ & $ 5.44 \pm 0.61$\\ 
TEETOA-G2p-G5 & $44.46 \pm 0.24$  & $39.08 \pm 0.27$ & $5.38 \pm 0.36$ & $ 5.13 \pm 0.60$\\ 
TEETOA-G3-G4  & $17.90 \pm 0.30$  & $18.82 \pm 0.24$ & $ -0.92 \pm 0.38$ & $ -0.86 \pm 0.62$\\ 
TEETOA-G3-G5  & $16.69 \pm 0.45$  & $17.09 \pm 0.48$ & $ -0.40 \pm 0.66$ & $ -1.17 \pm 0.61$\\ 
TEETOA-G4-G5  & $10.43 \pm 0.30$  & $10.47 \pm 0.24$ & $ -0.04 \pm 0.38$ & $ -0.31 \pm 0.61$\\ 
    \hline
    \end{tabular}
\begin{flushleft}\small
$^a$ In kcal/mol, errors are reported as 3 times the standard deviation. $^b$ This work. $^c$ from the differences of absolute binding free energies from Table \ref{tab:ATM-simulation-results}.\cite{Azimi2021SAMPL8}
\end{flushleft}
\end{table}

\subsection{Test of the Alignment Restraints}

As illustrated in Theory and Methods, the restraining potentials to keep the positions and orientations of the ligands in approximate alignment are designed to improve the convergence of the calculations without biasing the results. We tested this fact numerically on the RBFE calculation from G1 to G4 for the TEMOA host. The results, shown in Table \ref{tab:g1-g4-restraints}, largely confirm the theoretical expectations. The RBFE estimate obtained without alignment restraints (that is with only binding site restraints) agrees within statistical uncertainty with the estimate with full alignment restraints and with only positional restraints (that is by turning off orientation restraints). 

In addition to confirming the theory, this data shows that, despite the similarities between the G1 and G4 scaffolds (Figure \ref{fig:SAMPL8-systems}, a strict alignment between two ligands is not necessarily beneficial for the RBFE calculation. For example, the estimate obtained without angular restraints shows slightly smaller free energy changes with smaller statistical uncertainties in the two legs compared to the full restraints. This suggests that allowing for more flexibility can lower the free energy of the intermediate state in which the two ligands occupy simultaneously the binding cavity. 

The RBFE calculation without restraints, while also agreeing with the other estimates and with moderate uncertainty, suffered large entropic bottlenecks at the initial states of each leg (not shown) corresponding to the loss of translational and rotational degrees of freedom upon binding,\cite{pal2019perturbation} and required the use of the optimized softplus alchemical potential to reach convergence (see Computational Details). It can be thus concluded that some level of alignment restraints is beneficial to reach rapid convergence of the RBFE estimates with a general-purpose linear alchemical potential.

\begin{table}
\caption{Calculated binding free energy of TEMOA-G4 relative to TEMOA-G1 with various alignment restraints}
\label{tab:g1-g4-restraints}
\begin{tabular}{lccc}
Protocol                 & $\Delta G_1$      & $\Delta G2$       & $\Delta \Delta G^\circ_b$(RBFE) \\\hline
ABFE difference$^a$      &                   &                   & $ -1.92 \pm 0.42$  \\
RBFE full restraints$^a$ & $15.98 \pm 0.30$  & $18.45 \pm 0.39$  & $ -2.47 \pm 0.48$ \\
RBFE pos.\ restraints    & $15.26 \pm 0.20$  & $17.54 \pm 0.27$  & $ -2.28 \pm 0.34$  \\
RBFE no restraints       & $20.94 \pm 0.26$  & $23.39 \pm 0.30$  & $ -2.45 \pm 0.40$  \\
\noalign{\smallskip}\hline
\end{tabular}
\begin{flushleft}\small
$^a$ from Table \ref{tab:rbfe-sampl8}.
\end{flushleft}
\end{table}

\subsection{ER$\alpha$ Complexes}

The calculated RBFEs for all six pairs of the four ER$\alpha$ complexes (Figure \ref{fig:ERalpha-systems}) are shown in Table \ref{tab:rbfe-eralpha} compared to the corresponding values obtained by Wang et al.\cite{wang2017accurate} Given that were obtained with different force field models, the relative binding free energies calculated here for these systems are in good agreement with the literature values, which were themselves in reasonable agreement with experimental observations.\cite{wang2017accurate}  The values obtained here have similar uncertainties as the literature values and also display a good level of self-consistency with an average 3-point cycle closure error of $0.6$ kcal/mol, which is within statistical uncertainty.

One interesting and promising feature of the protein-ligand RBFEs obtained here is that the magnitudes of the free energy changes in each leg and their level of convergence do not appear to depend strongly on the difficulty of the alchemical transformation. For example, the large scaffold-hopping transformation from 2d to 3a which involves a ring expansion as well as the replacement and the displacement of a non-polar R-group with a polar one (Figure \ref{fig:ERalpha-systems}), presents leg free energies $\Delta G_1$ and $\Delta G_2$ which are only slightly larger in magnitude and with similar uncertainties that those of the 2d to 2e transformation which involves only a straightforward mutation of the three terminal atoms. 

\begin{table}[tb]
    \centering
    \caption{Relative binding free energies of the ER$\alpha$ complexes.}
    \label{tab:rbfe-eralpha}
    \begin{tabular}{lcccc}
Complex Pair       & $\Delta G_1$$^{a,b}$      & $\Delta G_2$$^{a,b}$     & $\Delta\Delta G^\circ_{\rm b}$(RBFE)$^{a,b}$ & $\Delta\Delta G^\circ_b$({\rm Wang et al.})$^{a,c}$ \\ \hline
2d-2e         & $12.67 \pm 0.28$  & $15.74 \pm 0.39$ & $ -3.06 \pm 0.48$ & $ -1.20 \pm 0.12$\\
2d-3a         & $17.01 \pm 0.39$  & $14.83 \pm 0.43$ & $  2.18 \pm 0.58$ & $  3.77 \pm 0.30$ \\
2d-3b         & $16.17 \pm 0.37$  & $16.30 \pm 0.40$ & $ -0.13 \pm 0.54$ & $  1.37 \pm 0.36$ \\
2e-3a         & $17.66 \pm 0.33$  & $13.35 \pm 0.34$ & $  4.31 \pm 0.48$ & $  5.26 \pm 0.30$ \\
2e-3b         & $13.87 \pm 0.27$  & $11.33 \pm 0.35$ & $  2.53 \pm 0.45$ & $  2.86 \pm 0.33$ \\
3a-3b         & $15.10 \pm 0.26$  & $17.70 \pm 0.32$ & $ -2.60 \pm 0.41$ & $ -2.36 \pm 0.33$ \\
    \hline
    \end{tabular}
\begin{flushleft}\small
$^a$ In kcal/mol, errors are reported as 3 times the standard deviation. $^b$ This work. $^c$ from reference \citenum{wang2017accurate}.
\end{flushleft}
\end{table}

\section{Discussion}

Binding free energy estimation remains one of the most challenging problem in modern computational chemistry. This is particularly so for the conformationally variable macromolecular targets relevant to structure-based drug discovery. Relative binding free energy (RBFE) methods based on alchemical transformations address some of these challenges by comparing directly pairs of ligands that share a similar binding mode thereby bypassing, for example, the modeling of the transition from the apo to the holo state of the receptor necessary for absolute binding free energy estimation (ABFE). However, alchemical RBFE methods based on conventional single- and dual-topology setups have been traditionally limited to simple R-group modifications, which considerably limits their range of applicability.

Recently, advanced alchemical scaffold-hopping methods have been introduced which allow for more complex changes in molecular topologies involving breaking and formation of covalent bonds. Wang et al.\cite{wang2017accurate} introduced a customized soft bond stretch potential to smoothly form and break bonds. More recently, Zou et al.\cite{zou2021scaffold} introduced a scheme based on auxiliary restraining potentials to achieve the same goal. Multi-site $\lambda$-dynamics has been presented as an alternative to scaffold hopping transformations involving ring opening and ring expansion/reduction.\cite{raman2020automated}
These advances have required a considerable amount of development and testing effort and are currently only available on commercial platforms--the Schr\"{o}dinger FEP+, the XTalPi XFEP, the BIOVIA Discovery Studio  platforms respectively. As academic efforts have lagged behind in this arena,\cite{jespers2019qligfep,vilseck2019overcoming,Rinikersubmitted} open-source software packages have not yet reached a comparable level of robustness for production applications.

Scaffold-hopping transformations are only one of the many challenges that prevent a widespread implementation and use of alchemical RBFE tools. Among others, single- and dual-topology RBFE implementations generally require the customization of energy and force routines, the design of pairwise soft-core potentials, and the treatment of dummy groups and their attachment to the core of the ligand. In addition, RBFE calculations generally require the mapping of the corresponding atoms of the ligands, the setup of separate simulations for the bound and unbound states, and the treatment of changes in net charge of the ligands.\cite{Mey2020Best,lee2020alchemical} The substantial intellectual and development efforts involved in the correct design, testing, and applications of alchemical RBFE software tools can be discouraging to many practitioners.

In this work, we have presented a streamlined RBFE protocol based on the Alchemical Transfer Method (ATM)\cite{wu2021alchemical} that promises to remove many of these obstacles while retaining a similar level of computational efficiency as the most advanced RBFE methods. The proposed ATM-RBFE protocol makes use of a single simulation system and does not require soft-core pair potentials and non-physical chemical topologies with dummy atoms. The method also does not require modifications of energy routines and can be easily implemented as a controlling routine on top of existing force routines of MD engines. The method has been implemented as open-source and freely available plugin of the OpenMM molecular simulation library.\cite{eastman2017openmm}

The method has its roots in the development of a statistical mechanics formulation of alchemical binding\cite{kilburg2018analytical} and the discovery of coordinate-based free energy perturbation schemes and novel alchemical potential functions capable of removing sampling bottlenecks along the alchemical thermodynamic path.\cite{pal2019perturbation} The resulting ATM method has been recently extended to the modeling of hydration\cite{khuttan2021single} and molecular binding\cite{wu2021alchemical,Azimi2021SAMPL8} with explicit solvation. In the present work, we combine ATM with a ligand alignment restraints protocol to model relative binding free energies. We tested the method on two stringent benchmarks that include scaffold-hopping and net charge modifications obtaining good agreement with ABFE calculations, literature values, and experimental observations.

\section{Conclusions}

We presented the theory and the implementation of the ATM-RBFE protocol for the streamlined calculation of relative binding free energies. We successfully tested the method on the SAMPL8 GDCC host-guest set and on a benchmark set of estrogen receptor $\alpha$ complexes including challenging scaffold-hopping transformations. 

\section{Acknowledgements}

We acknowledge support from the National Science Foundation (NSF
CAREER 1750511 to E.G.). Molecular simulations were conducted on the
Comet and Expanse GPU clusters at the San Diego Supercomputing Center supported by NSF XSEDE award TG-MCB150001.


\providecommand{\latin}[1]{#1}
\makeatletter
\providecommand{\doi}
  {\begingroup\let\do\@makeother\dospecials
  \catcode`\{=1 \catcode`\}=2 \doi@aux}
\providecommand{\doi@aux}[1]{\endgroup\texttt{#1}}
\makeatother
\providecommand*\mcitethebibliography{\thebibliography}
\csname @ifundefined\endcsname{endmcitethebibliography}
  {\let\endmcitethebibliography\endthebibliography}{}

\end{document}